\documentclass[12pt]{article}
\usepackage{graphicx}
\usepackage{amsmath}
\usepackage[symbol]{footmisc}

\renewcommand{\thefootnote}{\fnsymbol{footnote}}

\newcommand\pubnumber{DO-TH 22/27}
\newcommand\pubdate{November 23, 2022}

\def\institute{TU Dortmund University, Department of Physics, Otto-Hahn-Str.4, D-44221 Dortmund, Germany}
\def\Title#1{\begin{center} {\Large #1 } \end{center}}
\def\Author#1{\begin{center}{ \sc #1} \end{center}}
\def\Address#1{\begin{center}{ \it #1} \end{center}}

\newcommand\pubblock{\rightline{\begin{tabular}{l} \pubnumber\\
         \pubdate  \end{tabular}}}
\newenvironment{Abstract}{\begin{quotation}  }{\end{quotation}}
\newenvironment{Presented}{\begin{quotation} \begin{center}
             PRESENTED AT\end{center}\bigskip
      \begin{center}\begin{large}}{\end{large}\end{center} \end{quotation}}
\def\Acknowledgements{\bigskip  \bigskip \begin{center} \begin{large}
             \bf ACKNOWLEDGEMENTS \end{large}\end{center}}

\def\beq{\begin{equation}}
\def\eeq#1{\label{#1}\end{equation}}
\def\eeqn{\end{equation}}

\def\beqa{\begin{eqnarray}}
\def\eeqa#1{\label{#1}\end{eqnarray}}
\def\eeqan{\end{eqnarray}}

\let\bar=\overbar

\def\Dslash{\not{\hbox{\kern-4pt $D$}}}
\def\dslash{\not{\hbox{\kern-2pt $\del$}}}

\def\msb{{\bar{\ssstyle M \kern -1pt S}}}

\begin{document}
\begin{titlepage}
\pubblock

\vfill
\Title{Synergies of Drell-Yan, top and beauty in global SMEFT fits}
\vfill
\Author{Cornelius Grunwald, Gudrun Hiller, Kevin Kr\"oninger, Lara Nollen\footnote[1]{Speaker}}
\Address{\institute}
\vfill
\begin{Abstract}

We present a global fit of Drell-Yan, top-quark and beauty-physics data in the Standard Model Effective Field Theory framework using the Minimal Flavor Violation ansatz for the quark-flavor structure. The different energy scales are thereby connected by applying the renormalization group equation and matching onto the Weak Effective Theory in order to incorporate $b\to s$ FCNCs at low energies into the fit. We find that the combination of the different sectors greatly improves the bounds on the Wilson coefficients.

\end{Abstract}
\vfill
\begin{Presented}
$15^\mathrm{th}$ International Workshop on Top Quark Physics\\
Durham, UK, 4--9 September, 2022
\end{Presented}
\vfill
\end{titlepage}
\def\thefootnote{\fnsymbol{footnote}}
\setcounter{footnote}{0}

\section{Introduction}

Even though the Standard Model of particle physics (SM) describes a plethora of different observables with an impressive precision, the hunt for new physics (NP) is a very active field of research as there are still various observables that cannot be explained yet. Furthermore, the persistent anomalies observed in $B$-meson decays also hint towards a breakdown of the SM in precision observables connected to flavor.
However, going to much higher energy scales to directly search for the production of new particles at colliders does not seem feasible in the coming years, given the limitations of accelerator technology and the time it takes to build a new, larger collider. \\
In order to provide a complementary approach, we focus on the indirect search for physics beyond the Standard Model (BSM) using the Standard Model Effective Field Theory (SMEFT).
This framework allows for a largely model-independent examination of the effects of heavy NP in low-energy observables.
 We build upon the existing work in Ref.\,\cite{Bissmann:2020mfi} which we extend by including Drell-Yan data as well as by applying Minimal Flavor Violation (MFV) instead of the previously used top-philic approach.

\section{Standard Model Effective Field Theory}

The absence of a direct discovery of BSM physics at the LHC implies that new particles are either very weakly coupled to the SM or that the mass scale is well above the energy range currently probed at colliders.
The latter possibility entails that there exists a scale separation between the mass scale of SM particles and the NP scale $\Lambda$. With this assumption, the SMEFT Lagrangian can be constructed by systematically extending the SM Lagrangian with $d$-dimensional operators $O_i^{(d)}$ multiplied by a corresponding dimensionless Wilson coefficient $C_i^{(d)}$ which encodes the effects of possible BSM physics. The operators are constructed out of all the SM fields and are required to be invariant under the SM gauge symmetry SU(3)$_{{c}}\times$SU(2)$_{{L}}\times$U(1)$_{{Y}}$. In order to render the Lagrangian at mass dimension 4, each term in the Lagrangian is suppressed by $d-$4 powers of the high energy scale $\Lambda$, so that the SMEFT Lagrangian reads

\begin{equation}
      \mathcal{L}_{\text{SMEFT}}\,=\, \mathcal{L}_{\text{SM}}\,+ \,\sum_{d=5}^{\infty}\,\sum_i \: \frac{{C_i^{(d)}}}{\Lambda^{d-4}}{O_i^{(d)}} \: .
\end{equation}

In our analysis, we truncate this series at dimension 6 because higher dimensional operators are further suppressed by higher powers of $v^2/\Lambda^2$, with $v=246\,\text{GeV}$ the scale of electroweak symmetry breaking. At dimension 5, there is only the Weinberg operator, which is irrelevant to our fit as we only consider lepton number conserving processes. We work with the Warsaw basis \cite{Grzadkowski:2010es} and further assume all Wilson coefficients to be real valued for simplicity.

\section{Minimal Flavor Violation}

In order to reduce the number of degrees of freedom and impose a link between the different sectors, we use the MFV ansatz in the quark sector by imposing a $U(3)^3$ symmetry
\begin{equation}
      {G_F=U(3)_{Q_L}\times U(3)_{U_R} \times U(3)_{D_R}} \: .
\end{equation}
In this approach, the SM Yukawa matrices are treated as spurions. The SM fields and spurions transform under $G_F$ as
\begin{equation}
      Q_L \sim (3, 1, 1), \quad U_R \sim (1,3,1), \quad D_R \sim (1,1,3), \quad
      Y_u \sim (3,\bar{3}, 1), \quad Y_d \sim (3,1,\bar{3}).
\end{equation}
The SMEFT Lagrangian is required to be formally invariant under $G_F$, so that we can expand the flavor structure of the Wilson coefficients depending on the respective quark bilinear in terms of spurion insertions. For the example of two left handed quark doublets $\bar{Q} Q$, this expansion reads
\begin{equation}
      \quad C \sim a_1 {1}+a_2Y_uY_u^{\dagger}+a_3Y_dY_d^{\dagger}+ \mathcal{O}(Y_i)^4 \: .
\end{equation}
We retain terms up to order $Y_u^2$ and neglect all Yukawa couplings except for the~top-quark Yukawa $y_t$.
By performing a rotation to the mass basis, this expansion yields
\begin{equation}
      \begin{split}
            C_{ij}\: \bar{Q}_i\, Q_j \sim C \: \Biggl[ &\bar{u}_L
             \begin{pmatrix}
               a_1 & 0 & 0 \\
               0 & a_1 & 0 \\
               0 & 0 & a_1 +a_2\  y_t^2 \\
             \end{pmatrix}
             u_L + \\
             &\bar{d}_L
             \begin{pmatrix}
               a_1 + a_2\  \lvert V_{td} \rvert^2\ y_t^2 & a_2\ V_{td}^* V_{ts}\ y_t^2 & a_2\ V_{td}^* V_{tb}\ y_t^2 \\
               a_2\ V_{ts}^* V_{td}\ y_t^2 & a_1 + a_2\ \lvert V_{ts} \rvert^2\ y_t^2 & a_2\ V_{ts}^* V_{tb}\  y_t^2 \\
               a_2\ V_{tb}^* V_{td}\ y_t^2 & a_2\ V_{tb}^* V_{ts}\ y_t^2 & a_1 + a_2\ \lvert V_{tb} \rvert^2 \ y_t^2 \\
             \end{pmatrix}
             d_L \Biggr] \: .
      \end{split}
\end{equation}
We see that this imposes correlations among flavor entries and it allows for down-type FCNCs suppressed by the CKM elements $V_{ti}^* V_{tj}$ for a transition $d_i \to d_j$, similar to the SM.
This approach further solves the problem of unconstrained (flat) directions in the parameter space, which arise because usually, a linear combination of several operators contributes to an individual observable. By combining measurements of different linear combinations, this ambiguity can be resolved.

\section{Combined Fit}

\begin{figure}[h]
\centering
\includegraphics[height=3.6in]{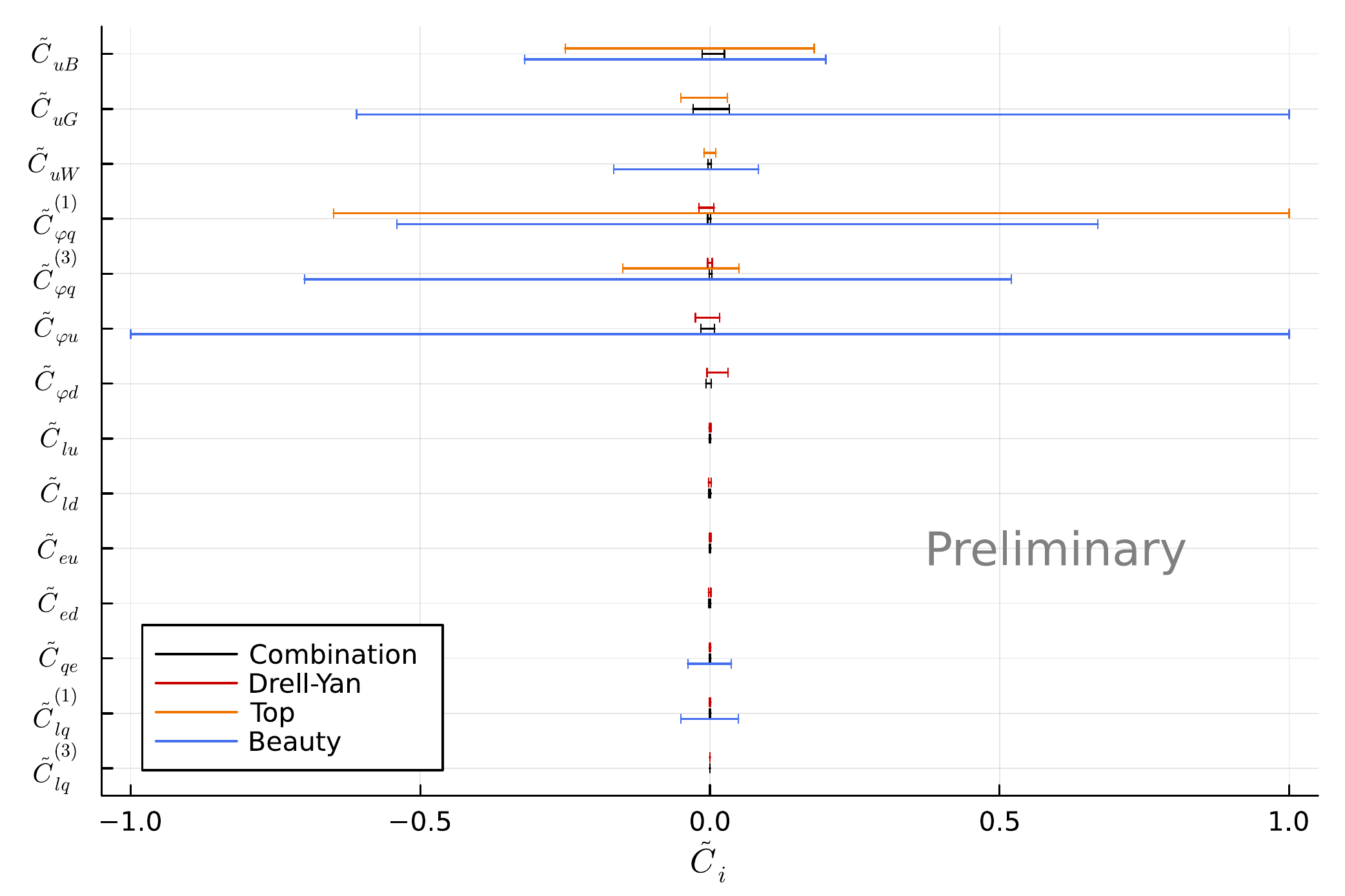}
\caption{Preliminary results of the global and individual fits. Shown are the 90\% credible intervals of the rescaled SMEFT Wilson coefficients assuming $\Lambda=1\,\text{TeV}$.}
\label{fig:results}
\end{figure}

In total, we consider 14 Wilson coefficients
($C_{uB}$, $C_{uG}$, $C_{uW}$, $C_{\varphi u}$, $C_{\varphi d}$, $C_{\varphi q}^{(1)}$, $C_{\varphi q}^{(3)}$,
$C_{lu}$, $C_{ld}$, $C_{eu}$, $C_{ed}$, $C_{lq}^{(1)}$, $C_{lq}^{(3)}$ and $C_{qe}$) and allow all operators to be present simultaneously \cite{Bissmann:2020mfi,Grundwald:2022}.
We include neutral-current as well as charged-current Drell-Yan measurements at the LHC, which are especially sensitive to the four-fermion operators due to their energy enhancement. In the top-quark sector, we include the total $t \bar t$ cross section, the total top-quark width, associated production cross sections ($t\bar t \gamma$ and $t\bar t Z$), as well as the helicity fractions. Additionally, we include $Z\to b\bar b$ transitions in order to constrain $Z$-vertex corrections. The $B$-meson observables considered in the analysis comprise cross sections, asymmetries and angular observables. They are included
by first performing the running of the Wilson coefficients with \textsf{\emph{wilson}} \cite{Aebischer:2018bkb} at one loop down to the scale $m_W$ of the $W$-boson mass and then matching to the Weak Effective Theory (WET) at one loop \cite{Dekens:2019ept}. Afterwards, the coefficients are evolved to the scale $m_b$ of the $b$-quark mass using the WET RGEs. \\
The fit is performed within Bayesian reasoning using EFTfitter \cite{Castro:2016jjv}. We rescale the Wilson coefficients as $\tilde C_i = \frac{v^2}{\Lambda^2}\, C_i$ and assume a flat prior distribution in the range of [-1,1] for each coefficient. The preliminary results of the combined fit as well as the fits of the individual data sets are shown in Fig \ref{fig:results}.
\\
We see that Drell-Yan measurements impose very tight limits on the semileptonic four-fermion Wilson coefficients of the order of $10^{-3}$, and they are moreover well suited to constrain quark vertex corrections with bounds of the order $10^{-2}$. They are, however, not sensitive to the dipole operators, which are best constrained by top-quark observables. The beauty observables impose limits on the up-type vertex corrections and dipole operators, which are however less tight compared to the other constraints as they mainly arise from the matching at one loop level. The tree-level contributions, in contrast, are suppressed by CKM-elements arising from the MFV ansatz. \\
In particular, we see that the limits of the combined fit can significantly improve the bounds compared to the individual fits, showing that a global fit is especially useful in the indirect search for BSM physics.

\section{Conclusion}
We explored the benefits from a combined fit of Drell-Yan, top-quark and beauty observables in the SMEFT framework based on the SU(2)$_L$-symmetry in the SMEFT and the MFV ansatz for the quark flavor. We observed that the bounds on the Wilson coefficients can be significantly improved by linking the different sectors.
In the future, additional measurements from the HL-LHC and Belle II will further improve the constraints. Another promising opportunity to enhance the constraints is at a future lepton collider.

\Acknowledgements
LN is very grateful to the organizers to be given the opportunity to present this work at the TOP 2022 conference. LN is supported by the doctoral scholarship program of the Studienstiftung des deutschen Volkes.


\begin{thebibliography}{99}


\bibitem{Bissmann:2020mfi}
S.~Bi\ss{}mann, C.~Grunwald, G.~Hiller and K.~Kr\"oninger,
JHEP \textbf{06} (2021), 010
doi:10.1007/JHEP06(2021)010
[arXiv:2012.10456 [hep-ph]].

\bibitem{Grzadkowski:2010es}
B.~Grzadkowski, M.~Iskrzynski, M.~Misiak and J.~Rosiek,
JHEP \textbf{10} (2010), 085
doi:10.1007/JHEP10(2010)085
[arXiv:1008.4884 [hep-ph]].

\bibitem{Aebischer:2018bkb}
J.~Aebischer, J.~Kumar and D.~M.~Straub,
Eur. Phys. J. C \textbf{78} (2018) no.12, 1026
doi:10.1140/epjc/s10052-018-6492-7
[arXiv:1804.05033 [hep-ph]].

\bibitem{Dekens:2019ept}
W.~Dekens and P.~Stoffer,
JHEP \textbf{10} (2019), 197
doi:10.1007/JHEP10(2019)197
[arXiv:1908.05295 [hep-ph]].

\bibitem{Castro:2016jjv}
N.~Castro, J.~Erdmann, C.~Grunwald, K.~Kr\"oninger and N.~A.~Rosien,
Eur. Phys. J. C \textbf{76} (2016) no.8, 432
doi:10.1140/epjc/s10052-016-4280-9
[arXiv:1605.05585].

\bibitem{Grundwald:2022}
C.~Grunwald, K.~Kr\"oninger G.~Hiller and L.~Nollen,
in preparation

\end{thebibliography}
\end{document}